\documentclass[epj]{svjour}
\usepackage{graphics}
\begin{document}

\authorrunning{Bogdanov}
\titlerunning{Neutron Star EoS Constraints with NICER}

\title{Prospects for Neutron Star Equation of State Constraints using ``Recycled'' Millisecond Pulsars}
\author{Slavko Bogdanov\inst{1}
}                     % Do not remove
\institute{Columbia Astrophysics Laboratory, Columbia University, 550 West 120th Street, New York, NY 10027, USA}
\date{Received: date / Revised version: date}
\abstract{``Recycled'' millisecond pulsars are a variety of rapidly-spinning
  neutron stars that typically show thermal X-ray radiation due to the heated
  surface of their magnetic polar caps. Detailed numerical modeling of
  the rotation-induced thermal X-ray pulsations observed from recycled
  millisecond pulsars, including all relevant relativistic and stellar
  atmospheric effects, has been identified as a promising approach
  towards an astrophysical determination of the true neutron star
  mass-radius relation, and by extension the state of cold matter at
  densities exceeding those of atomic nuclei. Herein, I review the
  basic model and methodology commonly used to extract information regarding
  neutron star structure from the pulsed X-ray radiation observed from
  millisecond pulsars. I also summarize the results of past X-ray
  observations of these objects and the prospects for precision
  neutron star mass-radius measurements with the upcoming Neutron Star
  Interior Composition Explorer (NICER) X-ray timing mission.}

\PACS{
      {04}{General relativity and gravitation}   \and
      {20}{Nuclear Physics}
     } % end of PACS codes
 %end of abstract
%
\maketitle
\section{Introduction}
\label{intro}
The state of cold matter at densities exceeding those of atomic nuclei
remains one of the principal outstanding problems in modern
physics. Neutron stars provide the only known setting in the Universe
where these physical conditions occur naturally.  Thermal X-ray
radiation from the physical surface of a neutron star can serve as a
powerful tool for probing the poorly understood behavior of the matter
in the ultra-dense stellar interior (see, e.g., \cite{Miller13} for a
comprehensive review). This is possible because of the unique mapping
between the pressure and density of neutron star matter and
the stellar radius and mass.  For neutron stars with thermal
radiation confined to a small fraction of the surface, realistic
modeling of the rotation-induced flux variations of the observed ``hot
spot'' radiation can, in principle, yield the true $M-R$ relation
\cite{Pavlov97} \cite{Bog07} \cite{Bog08}.  Such emission geometry is
observed in several varieties of neutron stars including
rotation-powered ``recycled'' millisecond pulsars (MSPs). Recycled
MSPs are a population of old neutron stars, characterized by rapid
rotation rates (hundreds of Hertz), exceptional rotational stability
and low magnetic fields ($\sim$10$^{8-9}$ G). It is commonly accepted
that these neutron stars are a product of low-mass X-ray binaries
\cite{Alp82}, acquiring their rapid spin rates via accretion of matter
and angular momentum. At the end of their spin-up phase, they are
reactivated (i.e., ``recycled'') as rotation-powered, radio-loud
pulsars, meaning that the observer radiation is generated at the
expense of the rotational kinetic energy of the neutron star.

 Over the past $\sim$15 years, extensive studies with \textit{Chandra}
 and \textit{XMM-Newton} have shown that many of these neutron stars
 are seen as X-ray sources due to their hot ($\sim$$10^6$ K) polar
 caps \cite{Zavlin06} \cite{Bog06a} \cite{Bog11} \cite{For14}. The
 inferred emitting areas indicate that this radiation is localized in
 regions on the stellar surface that are much smaller than the total
 surface area, but comparable in radius to what is expected for pulsar
 magnetic polar caps, $R_{pc}=(2\pi R/cP)^{1/2}$, where $R$ is the
 stellar radius and $P$ its spin period. This finding is consistent
 with pulsar electrodynamics models, which predict heating of the
 polar caps by a backflow of energetic particles along the open
 magnetic field lines \cite{Hard02}.  Sophisticated modeling of the
 X-ray spectra and pulse profiles (i.e. waveforms) of MSPs can offer a
 probe of the mass and radius of the star.  This approach, originally
 proposed by \cite{Pavlov97} in the context of ``recycled'' MSPs, can
 serve as a valuable probe of key NS properties that are inaccessible
 by other observational means.
 %Binary MSPs, in
 %particular, can offer very stringent EOS constraints via an
 %independent high-precision mass measurement from radio timing
 %combined with a $M/R$ measurement from X-ray observations.%

Below, I provide an overview of the standard approach used for
modeling the surface emission from neutron stars in general, and
rapidly spinning objects such as MSPs, in particular. I summarize the
application of this method in practice to existing data obtained with
\textit{XMM-Newton}. I conclude by describing the forthcoming Neutron
Star Interior Composition Explorer X-ray timing mission and the
precise measurement of neutron star structure it is expected to
achieve.

\section{Modeling the Surface Emission from Neutron Stars}
Conceptually, modeling the thermal radiation from a small region on a
slowly-spinning neutron star is fairly straightforward. The observational
characteristics of such a hot spot on a compact object were first
worked out by Pechenick, Ftaclas, and Cohen \cite{Pech83} and
variations and improvements of their formalism have subsequently been
presented in a number of other works \cite{Ftaclas86} \cite{Braje00}
\cite{Riff88} \cite{Miller98} \cite{Braje00} \cite{Belo02} \cite{Pou03}
\cite{Vii04}.

\begin{figure}
\rotatebox{270}{
  \resizebox{0.40\textwidth}{!}{ 
   \includegraphics{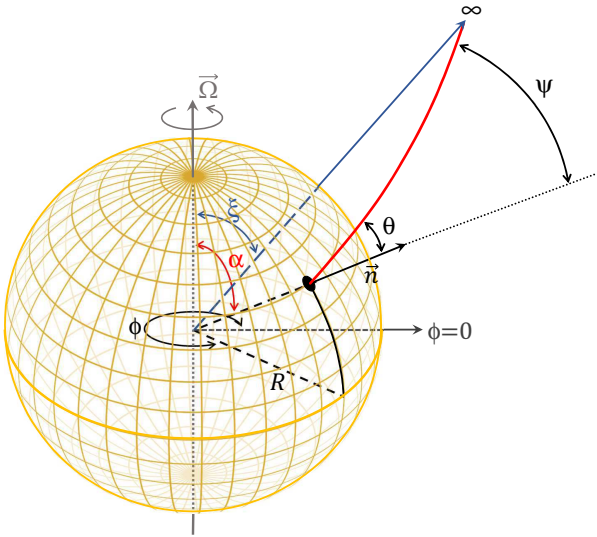}
}
}
\caption{\small{The geometry of a hot spot on the surface of a
    rotating neutron star. A non-radial photon emitted from the
    surface at an angle $\theta$ with respect to the local surface
    normal n is observed at infinity at an angle $\psi$ due to
    gravitational bending of light. The time-dependent position of the
    hot spot relative to the observer and the neutron star spin
    angular momentum vector is uniquely
    defined by the three angles $\alpha$, $\zeta$ and $\phi$ through
    equation (1).}}
\label{fig:nsdiagram}
\end{figure}

\subsection{System Geometry and Relativistic Effects}
In its simplest form, a model of a neutron star consists of a rotating
compact star of mass $M$, radius $R$, spin period $P$, and a single
small (point-like) hot spot, though it can be easily generalized to
include an arbitrary distribution of emitting surface elements.  The
hot spot is separated by an angle $\alpha$ from the rotation axis of
the star, which in turn is oriented at an angle $\zeta$ relative to
the line of sight to a distant observer (see the schematic
illustration in Figure 1). The position of a hot spot on the stellar
surface is defined by $\psi$, the angle between the normal to the
surface and the observer's line of sight:
\begin{equation}
\cos\psi(t)=\sin \alpha \sin \zeta \cos \phi (t) + \cos \alpha \cos \zeta
\end{equation}
The time-dependent angle $\phi(t)$ corresponds to the rotational phase
of the pulsar, with $\phi=0$ typically defined to be the time of
closest approach of the hot spot to the observer.

In the vicinity of a neutron star, gravity greatly affects the photons
as they propagate from the surface to a distant observed.  For
many practical purposes, the commonly used spherical Schwarzschild + Doppler
formalism provides a sufficiently accurate description of the
space-time near the stellar surface. It should be noted, however, that
for objects spinning at rates greater than $\sim$300 Hz, the rapid
rotation results in appreciable oblateness of the neutron star
although the external space-time is still well represented by the
Schwarzschild metric \cite{Cad07} \cite{Mor07}.
%causes the assumption of a
%spherical star to break down .
In such instances it is necessary to consider an oblate spheroid. For
MSPs with spin frequencies up to $\sim$800 Hz, the effect on the
stellar oblateness on the observed hot spot flux modulations can be as
high as 5--30\%, while the effect on the space-time quadrupole is
1-5\% \cite{Psaltis14b}. In this regime, the Hartle-Thorne metric
\cite{Hartle68} provides an accurate approximation of the space-time
\cite{Baub13}.  For the fastest spinning neutron stars (beyond
$\sim$1000 Hz), higher order space-time multipoles are non-negligible,
which necessitates numerically solving the field equations for a given
neutron star equation of state \cite{Cook94} \cite{Sterg95}.

As a photon climbs out of the deep gravitational potential, its
energy is diminished by $1+z_g=(1-R_S/R)^{-1/2}$, where $R_S=2GM/c^2$,
Additionaly, the trajectory of a photon emitted at an angle $\theta>0$
relative to the local radial direction is deflected, resulting in an
angle $\psi>\theta$ measured at infinity.  In Schwarzschild geometry,
the relation between these two quantities is expressed by the
elliptical integral \cite{Pech83}:
\begin{equation}
\psi= \int_{R}^{\infty}\frac{{\rm d}r}{r^2}\left[\frac{1}{b^2}-\frac{1}{r^2}\left(1-\frac{R_S}{r}\right)\right]^{-1/2}
\end{equation}
where 
\begin{equation}
b=\frac{R}{\sqrt{1-R_S/R}}\sin\theta
\end{equation}
is the impact parameter of a light ray originating from radius $R$ (at
the neutron star surface) that is emitted at an angle $\theta$. As the
use of the ray-tracing integral is computationally demanding, for many
applications it is more convenent to use the greatly simplified
approximate formula \cite{Belo02} \cite{Pou06}
\begin{equation}
\cos\psi\approx\frac{\cos\theta-R_S/R}{1-R_S/R}
\end{equation}
which can be used for $R > 2R_S$, where it achieves fractional errors
of only a few percent at the largest values of $\theta$. However, if
high accuracy is desired, the exact expression needs to be used.  Due
to the deflection of the photon paths by the immense gravitational
field, a larger fraction of the stellar surface is visible to an
observer at any instance. In the weak-field regime, the visibility
condition is simply $\cos\psi=\cos\theta>0$. In contrast, in the
presence of strong gravity regions on the far side of the neutron
star relative to the observer surface are viewable up to an angle
$\cos\psi_c$, corresponding to the maximum impact parameter $b_{\rm
  max}=R/\sqrt{1-R_S/R}\equiv R^{\infty}$, the so-called radius at
infinity.

For MSPs, the rapid motion of the neutron star surface induces a
substantial Doppler effect, parameterized through the familiar Doppler
factor
\begin{equation}
\eta=\frac{1}{\gamma(1-v/c \cos \xi)}
\end{equation}
where $\gamma=1/\sqrt{1-(v/c)^2}$, $v=2\pi R/P(1-R_S/R)^{-1/2}
\sin\alpha$ is the velocity of the emitting region as measured in the
inertial frame of the neutron star surface, and $\xi$ is the angle
between the directions of the velocity vector and towards the
observer.  This angle can be cast as a function of $\theta$, $\psi$,
$\zeta$, and $\phi$ \cite{Pou03} \cite{Vii04} as
\begin{equation}
\cos\xi=-\frac{\sin\theta}{\sin\psi}\sin \zeta \sin \phi
\end{equation}
A simplified approximate expression can be obtained by substituting
$\sin\theta/\sin\psi$ with $(1-R_S/R)^{1/2}$, its asymptotic value for
small angles.

\begin{figure}
 \resizebox{0.45\textwidth}{!}{
   \includegraphics{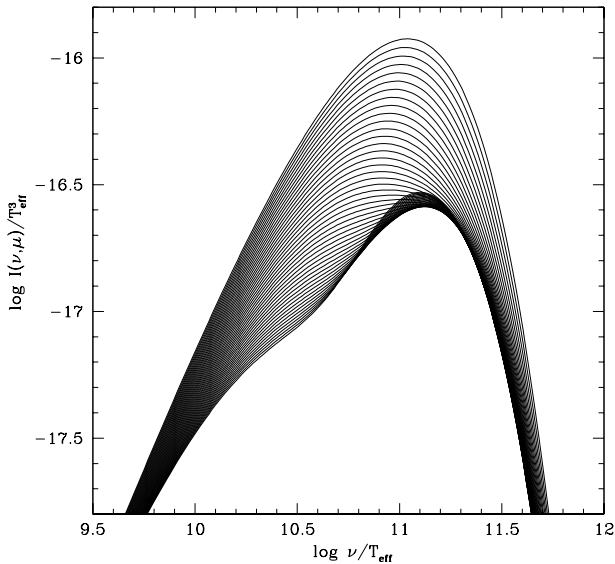}
}  
  \caption{The emergent intensity of a non-magnetic neutron star
    hydrogen atmosphere with an effective temperature of
    $2.1\times10^6$ Kelvin and a surface gravity of $2.4 \times
    10^{14}$ cm s$^{-2}$ (corresponding to a neutron star with mass
    1.4 $M_{\odot}$ and intrinsic radius $12$ km). The different spectra
    correspond to increasing emission angle with respect to the
    surface normal (from top to bottom, respectively) in logarithmic
    steps of 0.3 in $\cos\theta$. Note the shift in
    the peak of the spectrum towards lower energies in with increasing
    angle, in addition to the overall decline in intensity. This
    energy-dependent limb-darkening effect arises due to an
    temperature gradient within the atmosphere with temperature
    decreasing towards the surface combined with the fact that at
    larger viewing angles the observer sees less deep into the
    atmosphere.}
\label{fig:nsatmos}       % Give a unique label
\end{figure}

Photons emitted from the back side of the compact object as seen by
the observer, in addition to following a curved trajectory, have to
travel an additional distance compared to a photon emitted radially
from the near side. The time lag of the photon as recorded by an
observer at infinity is given by the elliptical integral \cite{Pech83}
\begin{equation}
\Delta t(b)=\frac{1}{c} \int_{R}^{\infty}\frac{{\rm d} r}{1-R_S/R} \Bigg\{\left[1-\frac{
b^2}{r^2}\left(1-\frac{R_S}{r}\right)\right]^{-1/2}-1\Bigg\}
\end{equation}
This time delay translates into a phase lag ($\Delta\phi$) of a photon
\begin{equation}
\Delta \phi =\frac{2 \pi}{P}\Delta t
\end{equation}
which produces the measured rotational phase $\phi_{\rm obs} =
\phi+\Delta \phi$ \cite{Vii04}.  For $R/R_S= 2.5$, the largest value
of $\Delta t$, obtained for light rays with maximum impact parameter
$b_{\rm max}=(1+z_g)R$, is $\approx$$60$ $\mu$s. These propagation time
differences amount to a few percent of the rotation period of a
typical MSP so they need to be taken into account when considering
 high-quality data.

The flux per unit frequency from a hot spot on a neutron star measured by a distant observed can be expressed as
\begin{equation}
F(\nu)=I(\nu){\rm d}\Omega
\end{equation}
where $I(\nu)$ is the intensity of the radiation as measured at infinity
and ${\rm d}\Omega$ is the apparent solid angle subtended by the hot
spot on the sky.  Transforming both quantities to the rest frame of the hot spot
yields
\begin{equation}
F(\nu)=(1-R_S/R)^{1/2}\eta^3 I'(\nu',\theta')\cos\theta'\frac{{\rm d} \cos\theta}{{\rm d} \cos\psi} \frac{{\rm d}S'}{D^2}
\end{equation}
Here, the variables marked with primes are measured in the rest frame
of the stellar surface \cite{Pou03}, with $\cos\theta'=\eta\cos\theta$
and ${\rm d}S\cos\theta={\rm d}S'\cos\theta'$.  $I'(\nu',\theta')$ is
the emergent radiation intensity, ${\rm d}S'$ is the emission area and $D$ is
the distance between the star and observer.  The three Doppler factors
are a result of the transformation of the intensity. An additional factor
is obtained upon integration over a frequency interval considering that ${\rm
  d}\nu=(1-R_S/R)^{1/2}\eta{\rm d}\nu'$.  Using equation (10), the
time-dependent flux observed from the rotating hot spot can be
determined for a given phase $\phi(t)$ (in the range 0 to $2\pi$)
using the relations between $\phi(t)$, $\theta$ and $\psi$ in equations
(1) and (2) and the appropriate emission model to compute $
I'(\nu',\theta')$, which is described next.

\begin{figure*}
\centering
\resizebox{0.95\textwidth}{!}{%
  \includegraphics{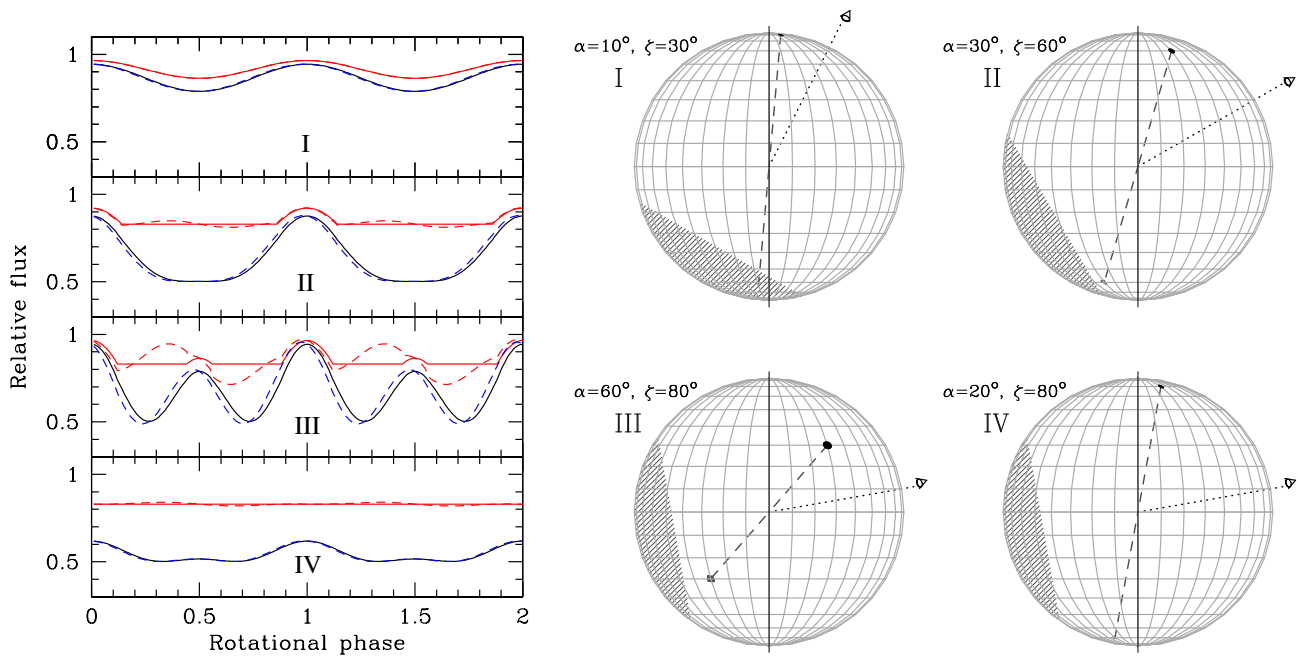}
}
%\vspace{5cm}       % Give the correct figure height in cm
\caption{(\textit{Left}) Representative model light curves for a
  rotating neutron star with $M=1.4$ M$_\odot$, $R=10$ km and two
  point-like antipodal hot spots for representative geometric
  configurations (\textit{right} panel).  The solid curves in each
  plot correspond to a H atmosphere (\textit{blue}) and isotropic
  blackbody emission (\textit{red}). The dashed lines show the
  effects of Doppler boosting and photon travel time delays for a spin
  frequency of 250 Hz.  (\textit{Right}) Orthographic map projection
  of the NS surface for the four pulse profiles (the roman numerals
  I--IV correspond to the lightcurves from top to bottom,
  respectively).  The dashed line is the axis connecting the two
  diametrically opposite hot spots while the dotted line is the
  direction to the observer. The hatched area corresponds to the
  portion of the star not visible to the observer. Due to
  gravitational bending of light, for a typical NS $\sim$80\% of the
  surface is visible at any given time.}
\label{fig:3}       % Give a unique label
\end{figure*}

\subsection{Neutron Star Atmosphere Emission}
The commonly accepted evolutionary scenario posits that MSPs acquire
their rapid spins due to accretion of matter and angular momentum in a
low-mass X-ray binary system \cite{Bhat91}.  Therefore, it is natural
to expect MSPs to posess a substantial atmospheric layer. Due to
gravitational settling, hydrogen is expected to surface within seconds
and dominate the surface emission. An optically-thick hydrogen
atmosphere of thickness $\sim$1 cm can by obtained with as little as
$10^{-20}$ M$_{\odot}$ of hydrogen.

Non-magnetic hydrogen atmosphere models applicable to MSPs have been
developed independently by different groups over the past 20 years
\cite{Zavlin96} \cite{Raj96} \cite{Heinke06} \cite{Haak12}.  They all
yield virtually identical results, with differences of only
$\sim$1\% around the peaks of the emergent spectra.  The models
consider a static, plane-parallel atmosphere that is in radiative
equilibrium, and composed of completely ionized hydrogen.  As
appropriate for MSPs, the surface is assumed to be weakly magnetized
($B \ll 10^{10}$ G), meaning that the effects of the magnetic field on
the opacity and equation of state of the atmosphere can be safely ignored.

Relative to a standard Planck spectrum, the radiation from a neutron
star atmosphere has peak emission that occurs at higher energies for
the same effective temperature and exhibits an overall flux depression,
which allows the conservation of bolometric flux \cite{Rom87}
\cite{Zavlin96}.  As a consequence, if modeled using a blackbody, a
neutron star covered by an atmosphere would be measured to have a
temperature much higher than its actual effective temperature, resulting in a grossly underestimated emitting area.  Furthermore,
the beaming pattern of the atmosphere is intrinsically non-uniform
with radiation intensity declining as the angle with respect to the
surface normal, resulting in the familiar limb-darkening effect (see
Figure 2).  In addition to a change in the total flux, there is a
significant shift in the peak energy of the spectrum.  For a blackbody
spectrum no such variations are expected. This implies that although
the emission spectrum is qualitatively similar to the case of a Planck
spectrum, the observed shape and photon energy dependence of the
rotation-induced pulsations of any localized emission on the surface
(such as hot spots seen from MSPs) will differ greatly.

Using the model ingredients described above, for MSPs, synthetic pulse
profiles can be generated by considering emission from two hot spots
(corresponding to the two dipole magnetic polar caps) as a function of
the neutron star rotational phase given a geometric configuration,
temperature, emission area, and input neutron star mass and radius
\cite{Bog07}. As seen Figure 3, the morphology of the flux
modulations are determined in large part by the geometric
configuration of the hot spot, neutron star spin axis and observer
system.  The choice of surface emission model is also a crucial factor
as apparent from the substantially larger amplitude of the H
atmosphere pulsations compared to a blackbody for the same assumed
parameters.

It is evident from equations (2) and (10) that the measured radiation
at infinity is highly sensitive to the choice of the neutron star
compactness, i.e. the mass-to-radius ratio ($M/R$). The impact of
$M/R$ on the rotation-induced X-ray modulations of a neutron star with
two diametrically opposite hot spots (corresponding to the two
magnetic poles of the star) covered with a non-magnetic neutron star
hydrogen atmosphere. A small increase in $M/R$ results in a pronounced
decrease in the amplitude of the pulsations (Figure 4), as a direct
result of the strong dependence of the magnitude of the bending of
light effect on $M/R$. Thus, as shown by \cite{Pavlov97} and
\cite{Zavlin98}, modeling of the pulsations of MSPs may can provide
constraints on $M$ and $R$.

\begin{figure}
%\centering
  %\vspace*{5cm}       % Give the correct figure height in cm
\resizebox{0.45\textwidth}{!}{
  \includegraphics{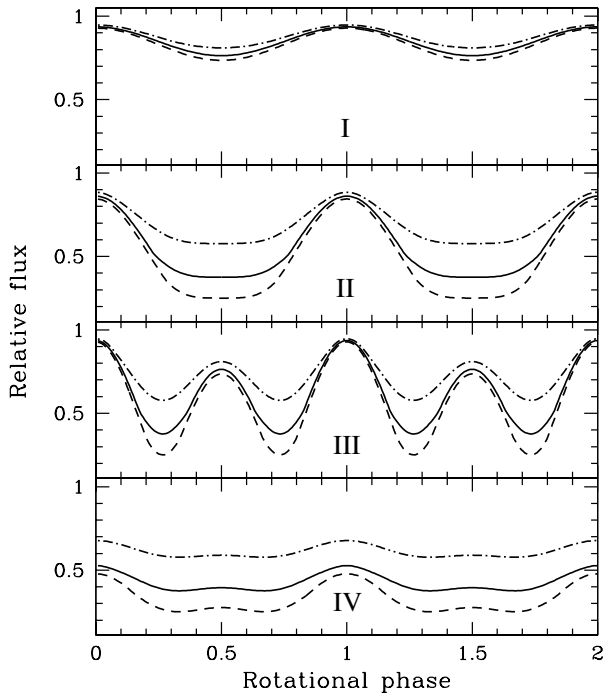}
}  
  \caption{Synthetic hydrogen atmosphere light curves for different
    stellar radii for a $1.4$ M$_\odot$ neutron star star. The lines
    correspond to stellar radii of $9$ km (\textit{dot-dashed}), $12$
    km (\textit{solid}), and $16$ km (\textit{dashed}). The angles
    $\alpha$ and $\zeta$ for each panel are assumed to have the same
    values as in Figure 2. Note the dramatic change in the amplitude
    of the rotation-induced flux variations as a function of stellar
    radius, caused by amplified bending of light effect for more
    compact stars. Two neutron star spin cycles are shown for
    clarity. Adapted from \cite{Bog07}.}
\label{fig:2}       % Give a unique label
\end{figure}

\section{Observational Results}
Rotation-powered MSPs were identified as pulsed X-ray sources by
Becker and Tr\"umper \cite{Beck93} in data from the \textit{ROSAT}
all-sky survey. The potential utillity of recycled MSPs as powerful
probes of the neutron star equation of state was first pointed out by
Pavlov and Zavlin \cite{Pavlov97} \cite{Zavlin98}, who used
\textit{ROSAT} data of the nearest known MSP, PSR J0437--4715
\cite{John93}, to demonstrate that a model of polar cap thermal
emission from a neutron star hydrogen atmosphere provides a good
description of the X-ray pulse profiles of this MSP, as well as to
place crude limits on the mass-radius relation.

\begin{figure}
 \resizebox{0.45\textwidth}{!}{
  \includegraphics{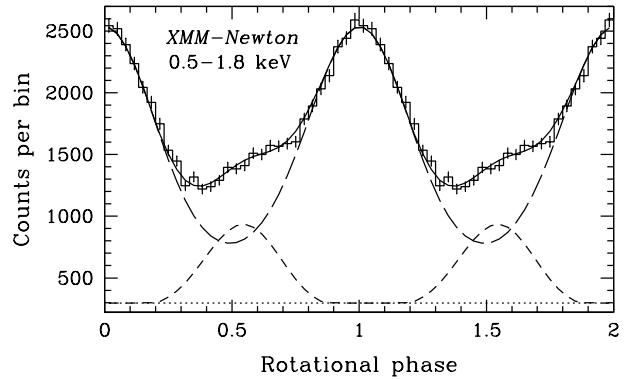}
}  
  \caption{The \textit{XMM-Newton} EPIC pn pulsations of the
    millisecond pulsar PSR J0437--4715 in the 0.5--1.8 keV energy band
    with the best fit model of a rotating neutron star with two
    X-ray-emitting polar caps covered by a non-magnetic hydrogen
    atmosphere (\textit{solid line}). The individual contributions
    from each hot spot are shown with the dashed lines. Note the
    offset from $\phi=0.5$ of the secondary pulse, indicating a
    displacement of the dipole field from the center of the star. The
    dotted line shows the background level (see \cite{Bog13} for
    further details).}
\label{fig:j0437}       % Give a unique label
\end{figure}

Prompted by this promising result,deep
\textit{XMM-Newton}\footnote{The {\it X-ray Multi Mirror-Newton} is an
  ESA science mission with instruments and contributions directly
  funded by ESA Member States and NASA.} European Photon Imaging
Camera (EPIC) pn observations of nearby MSPs were conducted
\cite{Bog08} \cite{Bog09} \cite{Bog13}. These efforts confirmed that a
non-magnetic hydrogen atmosphere can indeed reproduce the resulting
energy-dependent X-ray pulse profiles of the two closest known MSPs,
PSRs J0437--4715 and J0030+0451 (see Figures 4 and 5).  In contrast,
the large-amplitude pulsations are found to be incompatible with a
model that considers an isotropically-emitting Planck spectrum.
Furthermore, this modeling has already produced interesting
constraints on the allowed neutron star equation of state
(Figure~6). For PSR J0437--4715 (Figure 5), assuming $1.76$
M$_{\odot}$ (the current measurement from radio timing \cite{Verb08})
the stellar radius is constrained to be $R>11.1$ km (at 3$\sigma$
confidence; Bogdanov 2013), while for PSR J0030+0451 (Figure 6) the
best constraint is $R > 10.4$ (at 99.9\% confidence) assuming $1.4$
M$_{\odot}$ \cite{Bog09}. These limits are already inconsistent with
certain quark star and kaon condensate equations of state,
illustrating that this method represents a beneficial approach to
probing the neutron star EoS.

Their low inferred surface magnetic fields ($\sim$$10^{8-9}$ G), small
emitting areas ($\le$3 km radius), extraordinary rotational stability,
and dominant and steady non-transient surface emission make MSPs
fairly ``clean'' laboratories for studies of fundamental neutron star
physics.  As such, they can provide constraints on NS structure via
thermal pulse shape modeling that are complementary to those derived
from other approaches (e.g., using thermonuclear bursts from X-ray
binaries \cite{Ozel09} \cite{Steiner10} \cite{Sulei11} \cite{Pou14} and
spectroscopy of quiescent X-ray binaries \cite{Serv12} \cite{Guil13}
\cite{Heinke14}) and thus warrant extensive studies at X-ray energies.

One advantage of using recycled MSPs is the availability of (or the
possibility of obtaining) highly precise distance measurement from
very long baseline interferometry (VLBI) or high-precision radio
pulsar timing. The most notable example is PSR J0437--4715 for which
the parallax distance has been measured to within an unprecedented
$\pm$0.8\% ($156.3\pm1.3$ parsecs) \cite{Deller08}. This greatly
diminishes the uncertainty introduced in the emitting area, which (as
seen from equation 10) is strongly covariant with the distance between
the neutron star and the observer.

Perhaps more importantly, binary MSPs can offer particularly stringent
constraints on the equation of state via an independent high-precision
mass measurement from radio pulse timing combined with a $M/R$
measurement from X-ray observations. There is growing observational
evidence that MSPs are systematically more massive than the canonical
$1.4$ M$_{\odot}$, as expected for neutron stars spun-up by accretion;
this group includes the two most massive neutron stars known, PSRs
J1614--2230 and J0348+0432 with $\approx$2 $M_{\odot}$
\cite{Demorest10} \cite{Anton10}.  This places them in an interesting
region of the $M-R$ plane, away from the locus of model tracks around
$R=10$ km and $M=1.4$ M$_{\odot}$ (see Figure~7).

Moreover, binary MSPs permit an independent determination of the
observer's viewing angle ($\zeta$) of the neutron star since its value
is expected to coincide with the measurable orbital inclination due to
the expected alignment of the spin and orbital angular momentum during
the accreting spin-up phase. When combined with additional geometric
constraints, e.g., from modeling of \textit{Fermi} LAT $\gamma$-ray
pulsations \cite{Venter09}, this reduces the number of free model
parameters even further, thereby providing much more refined
bounds on the mass-radius relation.

\begin{figure}
  %\vspace*{5cm}       % Give the correct figure height in cm

 \resizebox{0.45\textwidth}{!}{
   \includegraphics{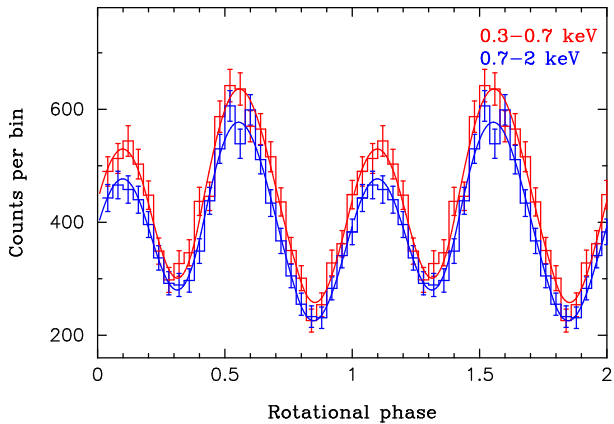}
}  
  \caption{The \textit{XMM-Newton} EPIC pn pulse profiles of the
    nearby isolated millisecond pulsar PSR J0030+0451 in the 0.3--0.7
    (red) and 0.7--2 (blue) keV ranges with the best fit model of a
    rapidly rotating neutron star with two H atmosphere hot polar caps
    \cite{Bog09}.}
\label{fig:j0030}       % Give a unique label
\end{figure}

\begin{figure}
  %\vspace*{5cm}       % Give the correct figure height in cm
\resizebox{0.45\textwidth}{!}{
  \includegraphics{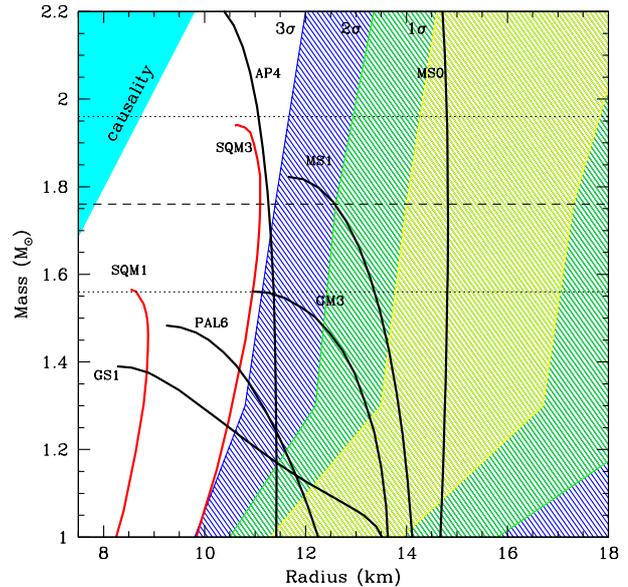}
}  
  \caption{The mass-radius plane for neutron stars showing the 1, 2,
    and 3$\sigma$ confidence contours (\textit{yellow},
    \textit{green}, and \textit{blue} hatched regions, respectively)
    for the millisecond pulsar PSR J0437--4715 based on a deep
    \textit{XMM-Newton} observation \cite{Bog13}. The solid lines are
    representative theoretical model tracks \cite{Lat01}. The
    horizontal lines show the pulsar mass measurement from radio
    timing (\textit{dashed line}) and the associated 1$\sigma$
    uncertainties (\textit{dotted lines}) \cite{Verb08}.}
\label{fig:mr}       % Give a unique label
\end{figure}

\section{The Neutron Star Interior Composition Explorer}
Existing X-ray data of MSPs are not of sufficient quality to provide
meaningful constraints on neutron star structure.  Nevertheless, the
have served to demonstrate that MSPs can serve as astrophysical
laboratories for studying ultra-dense matter. At present, further
improvements in neutron star mass-radius measurements of MSPs are
hindered by the design limitations of existing X-ray
observatories. Their potential utility makes MSPs obvious targets for
future X-ray observatories aimed at obtaining new and refining
existing measurements of the $M-R$ relation \cite{Motch13}.

Indeed, this has served as one of the principal science drivers for
the Neutron Star Interior Composition ExploreR (NICER) X-ray timing
instrument, currently scheduled for launch in late 2016.  NICER
is an approved NASA Explorer Mission of Opportunity that will be
deployed as an attached experiment on the International Space Station.
Its scientific payload is a non-imaging X-ray timing instrument that
is composed of an array of matching 56 X-ray concentrator and silicon
drift detector pairs sensitive in the 0.2--12 keV band. See Gendreau
et al. \cite{Gend12} for a more detailed overview of the design and
expected performance characteristics of NICER. Its unique combination of large
effective area (nearly 2000 cm$^{-2}$ at 1.5 keV), relatively low
background, and high precision timing capabilities ($\sim$100
nanoseconds absolute time resolution) is specifically tailored for
effective studies of the thermal X-ray pulsations from MSPs.

Within the nominal 18-month science mission, NICER will target the
nearest recycled MSPs, PSRs J0437--4715, J0030+0451, J2124--3358 and
perhaps others, in very long exposures ($\sim$1--1.5 Megaseconds),
which will produce $\sim$$1\times10^6$ source photons for each
target. This number of counts is sufficient to arrive at a $\sim$5\%
uncertainty in the measurement of the neutron star radius
\cite{Psaltis14}.  As demonstrated by \cite{Ozel10}, measuring the
mass-radius relation of several neutron stars to better than 10\%
would enable very strong limits on the allowed equation of state at
extreme densities \cite{Ozel10}. Therefore, observations of MSPs with
NICER hold the promise to produce a strong  empirical constraint
of the long sought-after pressure-density relation of cold
supra-nuclear matter, which would have profound implications for astrophysics
and nuclear physics alike.

% BibTeX users please use
 \bibliographystyle{plain}
 \bibliography{mybib}

\begin{thebibliography}{10}

\bibitem{Alp82}
M.~A. {Alpar}, A.~F. {Cheng}, M.~A. {Ruderman}, and J.~{Shaham}.
\newblock {A new class of radio pulsars}.
\newblock {\em Nature}, 300:728--730, December 1982.

\bibitem{Anton10}
J.~{Antoniadis}, P.~C.~C. {Freire}, N.~{Wex}, T.~M. {Tauris}, R.~S. {Lynch},
  M.~H. {van Kerkwijk}, M.~{Kramer}, C.~{Bassa}, V.~S. {Dhillon}, T.~{Driebe},
  J.~W.~T. {Hessels}, V.~M. {Kaspi}, V.~I. {Kondratiev}, N.~{Langer}, T.~R.
  {Marsh}, M.~A. {McLaughlin}, T.~T. {Pennucci}, S.~M. {Ransom}, I.~H.
  {Stairs}, J.~{van Leeuwen}, J.~P.~W. {Verbiest}, and D.~G. {Whelan}.
\newblock {A Massive Pulsar in a Compact Relativistic Binary}.
\newblock {\em Science}, 340:448, April 2013.

\bibitem{Baub13}
M.~{Baub{\"o}ck}, E.~{Berti}, D.~{Psaltis}, and F.~{{\"O}zel}.
\newblock {Relations between Neutron-star Parameters in the Hartle-Thorne
  Approximation}.
\newblock {\em ApJ}, 777:68, November 2013.

\bibitem{Beck93}
W.~{Becker} and J.~{Tr{\"u}mper}.
\newblock {Detection of pulsed X-rays from the binary millisecond pulsar J0437
  - 4715}.
\newblock {\em Nature}, 365:528--530, October 1993.

\bibitem{Belo02}
A.~M. {Beloborodov}.
\newblock {Gravitational Bending of Light Near Compact Objects}.
\newblock {\em ApJ}, 566:L85--L88, February 2002.

\bibitem{Bhat91}
D.~{Bhattacharya} and E.~P.~J. {van den Heuvel}.
\newblock {Formation and evolution of binary and millisecond radio pulsars}.
\newblock {\em PhR}, 203:1--124, 1991.

\bibitem{Bog13}
S.~{Bogdanov}.
\newblock {The Nearest Millisecond Pulsar Revisited with XMM-Newton: Improved
  Mass-radius Constraints for PSR J0437-4715}.
\newblock {\em ApJ}, 762:96, January 2013.

\bibitem{Bog09}
S.~{Bogdanov} and J.~E. {Grindlay}.
\newblock {Deep XMM-Newton Spectroscopic and Timing Observations of the
  Isolated Radio Millisecond Pulsar PSR J0030+0451}.
\newblock {\em ApJ}, 703:1557--1564, October 2009.

\bibitem{Bog06a}
S.~{Bogdanov}, J.~E. {Grindlay}, C.~O. {Heinke}, F.~{Camilo}, P.~C.~C.
  {Freire}, and W.~{Becker}.
\newblock {Chandra X-Ray Observations of 19 Millisecond Pulsars in the Globular
  Cluster 47 Tucanae}.
\newblock {\em ApJ}, 646:1104--1115, August 2006.

\bibitem{Bog08}
S.~{Bogdanov}, J.~E. {Grindlay}, and G.~B. {Rybicki}.
\newblock {Thermal X-Rays from Millisecond Pulsars: Constraining the
  Fundamental Properties of Neutron Stars}.
\newblock {\em ApJ}, 689:407--415, December 2008.

\bibitem{Bog07}
S.~{Bogdanov}, G.~B. {Rybicki}, and J.~E. {Grindlay}.
\newblock {Constraints on Neutron Star Properties from X-Ray Observations of
  Millisecond Pulsars}.
\newblock {\em ApJ}, 670:668--676, November 2007.

\bibitem{Bog11}
S.~{Bogdanov}, M.~{van den Berg}, M.~{Servillat}, C.~O. {Heinke}, J.~E.
  {Grindlay}, I.~H. {Stairs}, S.~M. {Ransom}, P.~C.~C. {Freire},
  S.~{B{\'e}gin}, and W.~{Becker}.
\newblock {Chandra X-ray Observations of 12 Millisecond Pulsars in the Globular
  Cluster M28}.
\newblock {\em ApJ}, 730:81, April 2011.

\bibitem{Braje00}
T.~M. {Braje}, R.~W. {Romani}, and K.~P. {Rauch}.
\newblock {Light Curves of Rapidly Rotating Neutron Stars}.
\newblock {\em ApJ}, 531:447--452, March 2000.

\bibitem{Cad07}
C.~{Cadeau}, S.~M. {Morsink}, D.~{Leahy}, and S.~S. {Campbell}.
\newblock {Light Curves for Rapidly Rotating Neutron Stars}.
\newblock {\em ApJ}, 654:458--469, January 2007.

\bibitem{Cook94}
G.~B. {Cook}, S.~L. {Shapiro}, and S.~A. {Teukolsky}.
\newblock {Rapidly rotating neutron stars in general relativity: Realistic
  equations of state}.
\newblock {\em ApJ}, 424:823--845, April 1994.

\bibitem{Deller08}
A.~T. {Deller}, J.~P.~W. {Verbiest}, S.~J. {Tingay}, and M.~{Bailes}.
\newblock {Extremely High Precision VLBI Astrometry of PSR J0437-4715 and
  Implications for Theories of Gravity}.
\newblock {\em ApJ}, 685:L67--L70, September 2008.

\bibitem{Demorest10}
P.~B. {Demorest}, T.~{Pennucci}, S.~M. {Ransom}, M.~S.~E. {Roberts}, and
  J.~W.~T. {Hessels}.
\newblock {A two-solar-mass neutron star measured using Shapiro delay}.
\newblock {\em Nature}, 467:1081--1083, October 2010.

\bibitem{For14}
L.~M. {Forestell}, C.~O. {Heinke}, H.~N. {Cohn}, P.~M. {Lugger}, G.~R.
  {Sivakoff}, S.~{Bogdanov}, A.~M. {Cool}, and J.~{Anderson}.
\newblock {A Chandra look at the X-ray faint millisecond pulsars in the
  globular cluster NGC 6752}.
\newblock {\em MNRAS}, 441:757--768, June 2014.

\bibitem{Ftaclas86}
C.~{Ftaclas}, M.~W. {Kearney}, and K.~{Pechenick}.
\newblock {Hot spots on neutron stars. II - The observer's sky}.
\newblock {\em ApJ}, 300:203--208, January 1986.

\bibitem{Gend12}
K.~C. {Gendreau}, Z.~{Arzoumanian}, and T.~{Okajima}.
\newblock {The Neutron star Interior Composition ExploreR (NICER): an Explorer
  mission of opportunity for soft x-ray timing spectroscopy}.
\newblock In {\em Society of Photo-Optical Instrumentation Engineers (SPIE)
  Conference Series}, volume 8443 of {\em Society of Photo-Optical
  Instrumentation Engineers (SPIE) Conference Series}, page~13, September 2012.

\bibitem{Guil13}
S.~{Guillot}, M.~{Servillat}, N.~A. {Webb}, and R.~E. {Rutledge}.
\newblock {Measurement of the Radius of Neutron Stars with High Signal-to-noise
  Quiescent Low-mass X-Ray Binaries in Globular Clusters}.
\newblock {\em ApJ}, 772:7, July 2013.

\bibitem{Haak12}
C.~B. {Haakonsen}, M.~L. {Turner}, N.~A. {Tacik}, and R.~E. {Rutledge}.
\newblock {The McGill Planar Hydrogen Atmosphere Code (McPHAC)}.
\newblock {\em ApJ}, 749:52, April 2012.

\bibitem{Hard02}
A.~K. {Harding} and A.~G. {Muslimov}.
\newblock {Pulsar Polar Cap Heating and Surface Thermal X-Ray Emission. II.
  Inverse Compton Radiation Pair Fronts}.
\newblock {\em ApJ}, 568:862--877, April 2002.

\bibitem{Hartle68}
J.~B. {Hartle} and K.~S. {Thorne}.
\newblock {Slowly Rotating Relativistic Stars. II. Models for Neutron Stars and
  Supermassive Stars}.
\newblock {\em ApJ}, 153:807, September 1968.

\bibitem{Heinke14}
C.~O. {Heinke}, H.~N. {Cohn}, P.~M. {Lugger}, N.~A. {Webb}, W.~C.~G. {Ho},
  J.~{Anderson}, S.~{Campana}, S.~{Bogdanov}, D.~{Haggard}, A.~M. {Cool}, and
  J.~E. {Grindlay}.
\newblock {Improved mass and radius constraints for quiescent neutron stars in
  {$\omega$} Cen and NGC 6397}.
\newblock {\em MNRAS}, 444:443--456, October 2014.

\bibitem{Heinke06}
C.~O. {Heinke}, G.~B. {Rybicki}, R.~{Narayan}, and J.~E. {Grindlay}.
\newblock {A Hydrogen Atmosphere Spectral Model Applied to the Neutron Star X7
  in the Globular Cluster 47 Tucanae}.
\newblock {\em ApJ}, 644:1090--1103, June 2006.

\bibitem{John93}
S.~{Johnston}, D.~R. {Lorimer}, P.~A. {Harrison}, M.~{Bailes}, A.~G. {Lyne},
  J.~F. {Bell}, V.~M. {Kaspi}, R.~N. {Manchester}, N.~{D'Amico}, and
  L.~{Nicastro}.
\newblock {Discovery of a very bright, nearby binary millisecond pulsar}.
\newblock {\em Nature}, 361:613--615, February 1993.

\bibitem{Lat01}
J.~M. {Lattimer} and M.~{Prakash}.
\newblock {Neutron Star Structure and the Equation of State}.
\newblock {\em ApJ}, 550:426--442, March 2001.

\bibitem{Miller13}
M.~C. {Miller}.
\newblock {Astrophysical Constraints on Dense Matter in Neutron Stars}.
\newblock {\em ArXiv e-prints}, November 2013.

\bibitem{Miller98}
M.~C. {Miller} and F.~K. {Lamb}.
\newblock {Bounds on the Compactness of Neutron Stars from Brightness
  Oscillations during X-Ray Bursts}.
\newblock {\em ApJ}, 499:L37--L40, May 1998.

\bibitem{Mor07}
S.~M. {Morsink}, D.~A. {Leahy}, C.~{Cadeau}, and J.~{Braga}.
\newblock {The Oblate Schwarzschild Approximation for Light Curves of Rapidly
  Rotating Neutron Stars}.
\newblock {\em ApJ}, 663:1244--1251, July 2007.

\bibitem{Motch13}
C.~{Motch}, J.~{Wilms}, D.~{Barret}, W.~{Becker}, S.~{Bogdanov}, L.~{Boirin},
  S.~{Corbel}, E.~{Cackett}, S.~{Campana}, D.~{de Martino}, F.~{Haberl},
  J.~{in't Zand}, M.~{M{\'e}ndez}, R.~{Mignani}, J.~{Miller}, M.~{Orio},
  D.~{Psaltis}, N.~{Rea}, J.~{Rodriguez}, A.~{Rozanska}, A.~{Schwope},
  A.~{Steiner}, N.~{Webb}, L.~{Zampieri}, and S.~{Zane}.
\newblock {The Hot and Energetic Universe: End points of stellar evolution}.
\newblock {\em ArXiv e-prints}, June 2013.

\bibitem{Ozel10}
F.~{{\"O}zel}, G.~{Baym}, and T.~{G{\"u}ver}.
\newblock {Astrophysical measurement of the equation of state of neutron star
  matter}.
\newblock {\em Phys Rev D}, 82(10):101301, November 2010.

\bibitem{Ozel09}
F.~{{\"O}zel} and D.~{Psaltis}.
\newblock {Reconstructing the neutron-star equation of state from astrophysical
  measurements}.
\newblock {\em Phys Rev D}, 80(10):103003, November 2009.

\bibitem{Pavlov97}
G.~G. {Pavlov} and V.~E. {Zavlin}.
\newblock {Mass-to-Radius Ratio for the Millisecond Pulsar J0437-4715}.
\newblock {\em ApJ}, 490:L91--L94, November 1997.

\bibitem{Pech83}
K.~R. {Pechenick}, C.~{Ftaclas}, and J.~M. {Cohen}.
\newblock {Hot spots on neutron stars - The near-field gravitational lens}.
\newblock {\em ApJ}, 274:846--857, November 1983.

\bibitem{Pou06}
J.~{Poutanen} and A.~M. {Beloborodov}.
\newblock {Pulse profiles of millisecond pulsars and their Fourier amplitudes}.
\newblock {\em MNRAS}, 373:836--844, December 2006.

\bibitem{Pou03}
J.~{Poutanen} and M.~{Gierli{\'n}ski}.
\newblock {On the nature of the X-ray emission from the accreting millisecond
  pulsar SAX J1808.4-3658}.
\newblock {\em MNRAS}, 343:1301--1311, August 2003.

\bibitem{Pou14}
J.~{Poutanen}, J.~{N{\"a}ttil{\"a}}, J.~J.~E. {Kajava}, O.-M. {Latvala}, D.~K.
  {Galloway}, E.~{Kuulkers}, and V.~F. {Suleimanov}.
\newblock {The effect of accretion on the measurement of neutron star mass and
  radius in the low-mass X-ray binary 4U 1608-52}.
\newblock {\em MNRAS}, 442:3777--3790, August 2014.

\bibitem{Psaltis14b}
D.~{Psaltis} and F.~{{\"O}zel}.
\newblock {Pulse Profiles from Spinning Neutron Stars in the Hartle-Thorne
  Approximation}.
\newblock {\em ApJ}, 792:87, September 2014.

\bibitem{Psaltis14}
D.~{Psaltis}, F.~{{\"O}zel}, and D.~{Chakrabarty}.
\newblock {Prospects for Measuring Neutron-star Masses and Radii with X-Ray
  Pulse Profile Modeling}.
\newblock {\em ApJ}, 787:136, June 2014.

\bibitem{Raj96}
M.~{Rajagopal} and R.~W. {Romani}.
\newblock {Model Atmospheres for Low-Field Neutron Stars}.
\newblock {\em ApJ}, 461:327, April 1996.

\bibitem{Riff88}
H.~{Riffert} and P.~{Meszaros}.
\newblock {Gravitational light bending near neutron stars. I - Emission from
  columns and hot spots}.
\newblock {\em ApJ}, 325:207--217, February 1988.

\bibitem{Rom87}
R.~W. {Romani}.
\newblock {Model atmospheres for cooling neutron stars}.
\newblock {\em ApJ}, 313:718--726, February 1987.

\bibitem{Serv12}
M.~{Servillat}, C.~O. {Heinke}, W.~C.~G. {Ho}, J.~E. {Grindlay}, J.~{Hong},
  M.~{van den Berg}, and S.~{Bogdanov}.
\newblock {Neutron star atmosphere composition: the quiescent, low-mass X-ray
  binary in the globular cluster M28}.
\newblock {\em MNRAS}, 423:1556--1561, June 2012.

\bibitem{Steiner10}
A.~W. {Steiner}, J.~M. {Lattimer}, and E.~F. {Brown}.
\newblock {The Equation of State from Observed Masses and Radii of Neutron
  Stars}.
\newblock {\em ApJ}, 722:33--54, October 2010.

\bibitem{Sterg95}
N.~{Stergioulas} and J.~L. {Friedman}.
\newblock {Comparing models of rapidly rotating relativistic stars constructed
  by two numerical methods}.
\newblock {\em ApJ}, 444:306--311, May 1995.

\bibitem{Sulei11}
V.~{Suleimanov}, J.~{Poutanen}, M.~{Revnivtsev}, and K.~{Werner}.
\newblock {A Neutron Star Stiff Equation of State Derived from Cooling Phases
  of the X-Ray Burster 4U 1724-307}.
\newblock {\em ApJ}, 742:122, December 2011.

\bibitem{Venter09}
C.~{Venter}, A.~K. {Harding}, and L.~{Guillemot}.
\newblock {Probing Millisecond Pulsar Emission Geometry Using Light Curves from
  the Fermi/Large Area Telescope}.
\newblock {\em ApJ}, 707:800--822, December 2009.

\bibitem{Verb08}
J.~P.~W. {Verbiest}, M.~{Bailes}, W.~{van Straten}, G.~B. {Hobbs}, R.~T.
  {Edwards}, R.~N. {Manchester}, N.~D.~R. {Bhat}, J.~M. {Sarkissian}, B.~A.
  {Jacoby}, and S.~R. {Kulkarni}.
\newblock {Precision Timing of PSR J0437-4715: An Accurate Pulsar Distance, a
  High Pulsar Mass, and a Limit on the Variation of Newton's Gravitational
  Constant}.
\newblock {\em ApJ}, 679:675--680, May 2008.

\bibitem{Vii04}
K.~{Viironen} and J.~{Poutanen}.
\newblock {Light curves and polarization of accretion- and nuclear-powered
  millisecond pulsars}.
\newblock {\em A\&A}, 426:985--997, November 2004.

\bibitem{Zavlin06}
V.~E. {Zavlin}.
\newblock {XMM-Newton Observations of Four Millisecond Pulsars}.
\newblock {\em ApJ}, 638:951--962, February 2006.

\bibitem{Zavlin98}
V.~E. {Zavlin} and G.~G. {Pavlov}.
\newblock {Soft X-rays from polar caps of the millisecond pulsar J0437-4715}.
\newblock {\em A\&A}, 329:583--598, January 1998.

\bibitem{Zavlin96}
V.~E. {Zavlin}, G.~G. {Pavlov}, and Y.~A. {Shibanov}.
\newblock {Model neutron star atmospheres with low magnetic fields. I.
  Atmospheres in radiative equilibrium.}
\newblock {\em A\&A}, 315:141--152, November 1996.

\end{thebibliography}


@ARTICLE{Alp82,
   author = {{Alpar}, M.~A. and {Cheng}, A.~F. and {Ruderman}, M.~A. and 
	{Shaham}, J.},
    title = "{A new class of radio pulsars}",
  journal = {Nature},
 keywords = {Pulsars, Radio Astronomy, Accretion Disks, Binary Stars, Gamma Rays, Stellar Mass Accretion, X Rays},
     year = 1982,
    month = dec,
   volume = 300,
    pages = {728-730},
      doi = {10.1038/300728a0},
   adsurl = {http://adsabs.harvard.edu/abs/1982Natur.300..728A},
  adsnote = {Provided by the SAO/NASA Astrophysics Data System}
}



@ARTICLE{Bog13,
   author = {{Bogdanov}, S.},
    title = "{The Nearest Millisecond Pulsar Revisited with XMM-Newton: Improved Mass-radius Constraints for PSR J0437-4715}",
  journal = {ApJ},
archivePrefix = "arXiv",
   eprint = {1211.6113},
 primaryClass = "astro-ph.HE",
 keywords = {pulsars: general, pulsars: individual: PSR J0437-4715, stars: neutron, X-rays: stars},
     year = 2013,
    month = jan,
   volume = 762,
      eid = {96},
    pages = {96},
      doi = {10.1088/0004-637X/762/2/96},
   adsurl = {http://adsabs.harvard.edu/abs/2013ApJ...762...96B},
  adsnote = {Provided by the SAO/NASA Astrophysics Data System}
}


@ARTICLE{2011ApJ...730...81B,
   author = {{Bogdanov}, S. and {van den Berg}, M. and {Servillat}, M. and 
	{Heinke}, C.~O. and {Grindlay}, J.~E. and {Stairs}, I.~H. and 
	{Ransom}, S.~M. and {Freire}, P.~C.~C. and {B{\'e}gin}, S. and 
	{Becker}, W.},
    title = "{Chandra X-ray Observations of 12 Millisecond Pulsars in the Globular Cluster M28}",
  journal = {ApJ},
archivePrefix = "arXiv",
   eprint = {1101.4944},
 primaryClass = "astro-ph.HE",
 keywords = {globular clusters: general, globular clusters: individual: NGC 6626, pulsars: general, pulsars: individual: PSR B1821-24 PSR J1824-2452H, stars: neutron, X-rays: stars},
     year = 2011,
    month = apr,
   volume = 730,
      eid = {81},
    pages = {81},
      doi = {10.1088/0004-637X/730/2/81},
   adsurl = {http://adsabs.harvard.edu/abs/2011ApJ...730...81B},
  adsnote = {Provided by the SAO/NASA Astrophysics Data System}
}


@ARTICLE{2008ApJ...689..407B,
   author = {{Bogdanov}, S. and {Grindlay}, J.~E. and {Rybicki}, G.~B.},
    title = "{Thermal X-Rays from Millisecond Pulsars: Constraining the Fundamental Properties of Neutron Stars}",
  journal = {ApJ},
archivePrefix = "arXiv",
   eprint = {0801.4030},
 keywords = {Stars: Pulsars: General, Stars: Pulsars: Individual: Alphanumeric: PSR J0030+0451, Stars: Pulsars: Individual: Alphanumeric: PSR J2124-3358, Stars: Neutron, X-Rays: Stars},
     year = 2008,
    month = dec,
   volume = 689,
    pages = {407-415},
      doi = {10.1086/592341},
   adsurl = {http://adsabs.harvard.edu/abs/2008ApJ...689..407B},
  adsnote = {Provided by the SAO/NASA Astrophysics Data System}
}


@ARTICLE{2009ApJ...703.1557B,
   author = {{Bogdanov}, S. and {Grindlay}, J.~E.},
    title = "{Deep XMM-Newton Spectroscopic and Timing Observations of the Isolated Radio Millisecond Pulsar PSR J0030+0451}",
  journal = {ApJ},
archivePrefix = "arXiv",
   eprint = {0908.1971},
 primaryClass = "astro-ph.HE",
 keywords = {pulsars: general, pulsars: individual: PSR J0030+0451, relativity, stars: neutron, X-rays: stars},
     year = 2009,
    month = oct,
   volume = 703,
    pages = {1557-1564},
      doi = {10.1088/0004-637X/703/2/1557},
   adsurl = {http://adsabs.harvard.edu/abs/2009ApJ...703.1557B},
  adsnote = {Provided by the SAO/NASA Astrophysics Data System}
}


@ARTICLE{2014MNRAS.441..757F,
   author = {{Forestell}, L.~M. and {Heinke}, C.~O. and {Cohn}, H.~N. and 
	{Lugger}, P.~M. and {Sivakoff}, G.~R. and {Bogdanov}, S. and 
	{Cool}, A.~M. and {Anderson}, J.},
    title = "{A Chandra look at the X-ray faint millisecond pulsars in the globular cluster NGC 6752}",
  journal = {MNRAS},
archivePrefix = "arXiv",
   eprint = {1403.4624},
 primaryClass = "astro-ph.HE",
 keywords = {stars: neutron, pulsars: general, globular clusters: individual: NGC 6752, X-rays: binaries},
     year = 2014,
    month = jun,
   volume = 441,
    pages = {757-768},
      doi = {10.1093/mnras/stu559},
   adsurl = {http://adsabs.harvard.edu/abs/2014MNRAS.441..757F},
  adsnote = {Provided by the SAO/NASA Astrophysics Data System}
}


@ARTICLE{2007ApJ...670..668B,
   author = {{Bogdanov}, S. and {Rybicki}, G.~B. and {Grindlay}, J.~E.},
    title = "{Constraints on Neutron Star Properties from X-Ray Observations of Millisecond Pulsars}",
  journal = {ApJ},
   eprint = {astro-ph/0612791},
 keywords = {Gravitation, Stars: Pulsars: General, Stars: Pulsars: Individual: Alphanumeric: PSR J0437-4715, Relativity, Stars: Neutron, X-Rays: Stars},
     year = 2007,
    month = nov,
   volume = 670,
    pages = {668-676},
      doi = {10.1086/520793},
   adsurl = {http://adsabs.harvard.edu/abs/2007ApJ...670..668B},
  adsnote = {Provided by the SAO/NASA Astrophysics Data System}
}


@ARTICLE{2006ApJ...646.1104B,
   author = {{Bogdanov}, S. and {Grindlay}, J.~E. and {Heinke}, C.~O. and 
	{Camilo}, F. and {Freire}, P.~C.~C. and {Becker}, W.},
    title = "{Chandra X-Ray Observations of 19 Millisecond Pulsars in the Globular Cluster 47 Tucanae}",
  journal = {ApJ},
   eprint = {astro-ph/0604318},
 keywords = {Galaxy: Globular Clusters: General, Galaxy: Globular Clusters: Individual: Name: 47 Tucanae, Stars: Pulsars: General, Stars: Neutron, X-Rays: Stars},
     year = 2006,
    month = aug,
   volume = 646,
    pages = {1104-1115},
      doi = {10.1086/505133},
   adsurl = {http://adsabs.harvard.edu/abs/2006ApJ...646.1104B},
  adsnote = {Provided by the SAO/NASA Astrophysics Data System}
}


@ARTICLE{2006ApJ...648L..55B,
   author = {{Bogdanov}, S. and {Grindlay}, J.~E. and {Rybicki}, G.~B.},
    title = "{X-Rays from Radio Millisecond Pulsars: Comptonized Thermal Radiation}",
  journal = {ApJ},
   eprint = {astro-ph/0605273},
 keywords = {Stars: Pulsars: General, Stars: Pulsars: Individual: Alphanumeric: PSR J0437-4715, Stars: Neutron, X-Rays: Stars},
     year = 2006,
    month = sep,
   volume = 648,
    pages = {L55-L58},
      doi = {10.1086/507827},
   adsurl = {http://adsabs.harvard.edu/abs/2006ApJ...648L..55B},
  adsnote = {Provided by the SAO/NASA Astrophysics Data System}
}


@ARTICLE{2006ApJ...638..951Z,
   author = {{Zavlin}, V.~E.},
    title = "{XMM-Newton Observations of Four Millisecond Pulsars}",
  journal = {ApJ},
   eprint = {astro-ph/0507235},
 keywords = {Stars: Pulsars: Individual: Alphanumeric: PSR J0034-0534, Stars: Pulsars: Individual: Alphanumeric: PSR J0437-4715, pulsars: individual (PSR J1024-0719), Stars: Pulsars: Individual: Alphanumeric: PSR J2124-3358, Stars: Neutron, X-Rays: Stars},
     year = 2006,
    month = feb,
   volume = 638,
    pages = {951-962},
      doi = {10.1086/449308},
   adsurl = {http://adsabs.harvard.edu/abs/2006ApJ...638..951Z},
  adsnote = {Provided by the SAO/NASA Astrophysics Data System}
}


@ARTICLE{1996A&A...315..141Z,
   author = {{Zavlin}, V.~E. and {Pavlov}, G.~G. and {Shibanov}, Y.~A.},
    title = "{Model neutron star atmospheres with low magnetic fields. I. Atmospheres in radiative equilibrium.}",
  journal = {A&A},
   eprint = {astro-ph/9604072},
 keywords = {STARS: NEUTRON, DENSE MATTER, MAGNETIC FIELDS, RADIATIVE TRANSFER, RADIATION MECHANISMS: THERMAL},
     year = 1996,
    month = nov,
   volume = 315,
    pages = {141-152},
   adsurl = {http://adsabs.harvard.edu/abs/1996A%26A...315..141Z},
  adsnote = {Provided by the SAO/NASA Astrophysics Data System}
}


@ARTICLE{2002ApJ...569..894Z,
   author = {{Zavlin}, V.~E. and {Pavlov}, G.~G. and {Sanwal}, D. and {Manchester}, R.~N. and 
	{Tr{\"u}mper}, J. and {Halpern}, J.~P. and {Becker}, W.},
    title = "{X-Radiation from the Millisecond Pulsar J0437-4715}",
  journal = {ApJ},
   eprint = {astro-ph/0112544},
 keywords = {Stars: Pulsars: Individual: Alphanumeric: PSR J0437-4715, Stars: Neutron, X-Rays: Stars},
     year = 2002,
    month = apr,
   volume = 569,
    pages = {894-902},
      doi = {10.1086/339351},
   adsurl = {http://adsabs.harvard.edu/abs/2002ApJ...569..894Z},
  adsnote = {Provided by the SAO/NASA Astrophysics Data System}
}


@ARTICLE{1998A&A...329..583Z,
   author = {{Zavlin}, V.~E. and {Pavlov}, G.~G.},
    title = "{Soft X-rays from polar caps of the millisecond pulsar J0437-4715}",
  journal = {A&A},
   eprint = {astro-ph/9708101},
 keywords = {STARS: NEUTRON, X-RAY: STARS, PULSARS: INDIVIDUAL: PSR J0437-4715},
     year = 1998,
    month = jan,
   volume = 329,
    pages = {583-598},
   adsurl = {http://adsabs.harvard.edu/abs/1998A%26A...329..583Z},
  adsnote = {Provided by the SAO/NASA Astrophysics Data System}
}


@ARTICLE{2004A&A...426..985V,
   author = {{Viironen}, K. and {Poutanen}, J.},
    title = "{Light curves and polarization of accretion- and nuclear-powered millisecond pulsars}",
  journal = {A&A},
   eprint = {astro-ph/0408250},
 keywords = {methods: numerical, polarization, stars: pulsars: general, stars: neutron, stars: oscillations, X-rays: binaries},
     year = 2004,
    month = nov,
   volume = 426,
    pages = {985-997},
      doi = {10.1051/0004-6361:20041084},
   adsurl = {http://adsabs.harvard.edu/abs/2004A%26A...426..985V},
  adsnote = {Provided by the SAO/NASA Astrophysics Data System}
}


@ARTICLE{1987ApJ...313..718R,
   author = {{Romani}, R.~W.},
    title = "{Model atmospheres for cooling neutron stars}",
  journal = {ApJ},
 keywords = {Atmospheric Models, Cooling, Neutron Stars, Stellar Atmospheres, Black Body Radiation, Gravitational Effects, Stellar Evolution, Surface Temperature, Thermodynamic Equilibrium, X Ray Sources},
     year = 1987,
    month = feb,
   volume = 313,
    pages = {718-726},
      doi = {10.1086/165010},
   adsurl = {http://adsabs.harvard.edu/abs/1987ApJ...313..718R},
  adsnote = {Provided by the SAO/NASA Astrophysics Data System}
}


@ARTICLE{2003MNRAS.343.1301P,
   author = {{Poutanen}, J. and {Gierli{\'n}ski}, M.},
    title = "{On the nature of the X-ray emission from the accreting millisecond pulsar SAX J1808.4-3658}",
  journal = {MNRAS},
   eprint = {astro-ph/0303084},
 keywords = {accretion, accretion discs, methods: data analysis, pulsars: individual: SAX J1808.4-3658, X-rays: binaries},
     year = 2003,
    month = aug,
   volume = 343,
    pages = {1301-1311},
      doi = {10.1046/j.1365-8711.2003.06773.x},
   adsurl = {http://adsabs.harvard.edu/abs/2003MNRAS.343.1301P},
  adsnote = {Provided by the SAO/NASA Astrophysics Data System}
}


@ARTICLE{1997ApJ...490L..91P,
   author = {{Pavlov}, G.~G. and {Zavlin}, V.~E.},
    title = "{Mass-to-Radius Ratio for the Millisecond Pulsar J0437-4715}",
  journal = {ApJ},
   eprint = {astro-ph/9709255},
 keywords = {STARS: PULSARS: INDIVIDUAL ALPHANUMERIC: PSR J0437-, 715, STARS: NEUTRON, X-RAYS: STARS, Stars: Pulsars: Individual: Alphanumeric: PSR J0437-4715, Stars: Neutron, X-Rays: Stars},
     year = 1997,
    month = nov,
   volume = 490,
    pages = {L91-L94},
      doi = {10.1086/311007},
   adsurl = {http://adsabs.harvard.edu/abs/1997ApJ...490L..91P},
  adsnote = {Provided by the SAO/NASA Astrophysics Data System}
}


@ARTICLE{2007ApJ...663.1244M,
   author = {{Morsink}, S.~M. and {Leahy}, D.~A. and {Cadeau}, C. and {Braga}, J.
	},
    title = "{The Oblate Schwarzschild Approximation for Light Curves of Rapidly Rotating Neutron Stars}",
  journal = {ApJ},
   eprint = {astro-ph/0703123},
 keywords = {Stars: Pulsars: General, Relativity, Stars: Neutron, Stars: Rotation, X-Rays: Binaries},
     year = 2007,
    month = jul,
   volume = 663,
    pages = {1244-1251},
      doi = {10.1086/518648},
   adsurl = {http://adsabs.harvard.edu/abs/2007ApJ...663.1244M},
  adsnote = {Provided by the SAO/NASA Astrophysics Data System}
}


@ARTICLE{2001ApJ...550..426L,
   author = {{Lattimer}, J.~M. and {Prakash}, M.},
    title = "{Neutron Star Structure and the Equation of State}",
  journal = {ApJ},
   eprint = {astro-ph/0002232},
 keywords = {Equation of State, Stars: Interiors, Stars: Neutron},
     year = 2001,
    month = mar,
   volume = 550,
    pages = {426-442},
      doi = {10.1086/319702},
   adsurl = {http://adsabs.harvard.edu/abs/2001ApJ...550..426L},
  adsnote = {Provided by the SAO/NASA Astrophysics Data System}
}



@ARTICLE{2002ApJ...568..862H,
   author = {{Harding}, A.~K. and {Muslimov}, A.~G.},
    title = "{Pulsar Polar Cap Heating and Surface Thermal X-Ray Emission. II. Inverse Compton Radiation Pair Fronts}",
  journal = {ApJ},
   eprint = {astro-ph/0112392},
 keywords = {Stars: Pulsars: General, Radiation Mechanisms: Nonthermal, Relativity, Stars: Neutron, X-Rays: Stars},
     year = 2002,
    month = apr,
   volume = 568,
    pages = {862-877},
      doi = {10.1086/338985},
   adsurl = {http://adsabs.harvard.edu/abs/2002ApJ...568..862H},
  adsnote = {Provided by the SAO/NASA Astrophysics Data System}
}


@ARTICLE{2007ApJ...654..458C,
   author = {{Cadeau}, C. and {Morsink}, S.~M. and {Leahy}, D. and {Campbell}, S.~S.
	},
    title = "{Light Curves for Rapidly Rotating Neutron Stars}",
  journal = {ApJ},
   eprint = {astro-ph/0609325},
 keywords = {Stars: Pulsars: General, Relativity, Stars: Neutron, Stars: Rotation},
     year = 2007,
    month = jan,
   volume = 654,
    pages = {458-469},
      doi = {10.1086/509103},
   adsurl = {http://adsabs.harvard.edu/abs/2007ApJ...654..458C},
  adsnote = {Provided by the SAO/NASA Astrophysics Data System}
}
%\begin{thebibliography}{}

%  Author, Journal \textbf{Volume}, (year) page numbers.
%Author, \textit{Book title} (Publisher, place year) page numbers
% etc
%\end{thebibliography}

\end{document}